\documentclass[wcp]{jmlr}

\usepackage{longtable}% for long tables
\usepackage{enumerate}
\usepackage{graphicx}
\usepackage{url}
\usepackage{booktabs}
\makeatletter
\let\Ginclude@graphics\@org@Ginclude@graphics 
\makeatother

\jmlrvolume{}
\jmlryear{2021}
\jmlrworkshop{ACML 2021}

\title[TSEGAN]{Time-domain Speech Enhancement with Generative Adversarial Learning}

  \author{\Name{Feiyang Xiao} \Email{xiaofeiyang128@gmail.com} \\
  \Name{Jian Guan}\footnotemark[1] \Email{j.guan@hrbeu.edu.cn}\\
  \addr Group of Intelligent Signal Processing, College of Computer Science and Technology \\Harbin Engineering University, Harbin, 150001, China (Corresponding Author is Jian Guan)\\
  \Name{Qiuqiang Kong} \Email{kongqiuqiang@bytedance.com}\\
  \addr ByteDance, Shanghai, 200233, China
  \AND
  \Name{Wenwu Wang} \Email{w.wang@surrey.ac.uk}\\
  \addr Centre for Vision Speech and Signal Processing, University of Surrey, Guildford, GU2 7XH, UK
  }

\begin{document}

\maketitle

\begin{abstract}
Speech enhancement aims to obtain speech signals of high intelligibility and quality from noisy speech. Recent work has demonstrated the excellent performance of time-domain deep learning methods, such as Conv-TasNet. However, these methods can be degraded by the arbitrary scales of the waveform induced by the scale-invariant signal-to-noise ratio (SI-SNR) loss. 
This paper proposes a new framework called Time-domain Speech Enhancement Generative Adversarial Network (TSEGAN), which is an extension of the generative adversarial network (GAN) in time-domain with metric evaluation to mitigate the scaling problem, and provide model training stability, thus achieving performance improvement.
In addition, we provide a new method based on objective function mapping for the theoretical analysis of the performance of Metric GAN, and explain why it is better than the Wasserstein GAN. 
Experiments conducted demonstrate the effectiveness of our proposed method, and illustrate the advantage of Metric GAN.
The source code  will be available at \url{https://github.com/LittleFlyingSheep/TSEGAN}.
\end{abstract}

\section{Introduction}
Speech enhancement aims to remove noise or interference from noisy speech in order to reconstruct the clean speech \citep{2,23,24,choi2018phase,xu2014regression}. It has widespread applications in e.g., human computer interaction, robust speech recognition \citep{du2020double}, hearing aids \citep{fedorov2020tinylstms}, and cochlear implants \citep{yang2005spectral}.

Recently, multi-head self-attention (MHSA) was shown to offer excellent performance in speech enhancement for unknown speakers \citep{2}. In \cite{23}, speech enhancement is achieved with speech presence probability estimation, showing good performance in terms of subjective evaluation metrics. In \cite{24}, gated convolutional recurrent networks are introduced for speech enhancement with phase estimation for the speech source to be recovered.

These methods all work in the time-frequency (T-F) domain, achieved by using short-time Fourier transform (STFT). However, the estimation of the phase information may not be accurate \citep{choi2018phase}. On the other hand, some studies even do not estimate the phase of the sources, but directly use the phase of the speech mixture~\cite{xu2014regression}. The inaccurate phase information can lead to possible errors in the enhanced speech when it is converted to the time domain with inverse STFT (iSTFT).
In recent studies, such as Conv-TasNet \citep{7,7-2,9}, the feature representation of speech is learned in a latent space in the time domain with a loss function based on scale invariant signal to noise ratio (SI-SNR), and is shown to offer advantages for speech enhancement \citep{7-2,9}. 
However, due to the use of  SI-SNR, the dynamic range of the waveform may be arbitrarily scaled in Conv-TasNet, which can potentially lead to instability in model training \citep{le2019sdr,21}. 

Recently, GAN models are shown to boost the generalization performance, and improve the quality of enhanced speech \citep{10,14}, as shown in the speech enhancement generative adversarial network (SEGAN) where conditional GAN is used for speech enhancement. Although SEGAN achieves good performance measured in subjective metrics, the performance measured via objective metrics such as SNR tends to be degraded, due to the vanishing gradient problem with the conditional GAN loss during training \citep{10}.

To address this problem, the Wasserstein distance \citep{12,13,17} has been introduced to improve the conditional GAN loss, resulting in the Wasserstein GAN (WGAN) method that achieves better objective performance than SEGAN \citep{14,15}. The WGAN method is further improved in \cite{16} by employing metric evaluation in the conditional GAN loss, and leads to the Metric GAN method, which outperforms WGAN based methods for speech enhancement. However, there is no existing explanation why Metric GAN is superior to WGAN. Note that, all these GAN methods work in the T-F domain.

In this paper,  we present a novel method called Time-domain Speech Enhancement GAN (TSEGAN), where we use TasNet (i.e., a version of Conv-TasNet without using the SI-SNR loss) as a generator to obtain the latent space representation of the input noisy speech, and design a metric evaluation discriminator to optimize the generator. Our contributions can be summarized as follows:
\begin{itemize}
\item We show that the use of SI-SNR loss can lead to 
unstable training in speech enhancement.
To address this issue, we incorporate metric evaluation in the discriminator to help mitigate performance degradation caused by the SI-SNR loss.
\item We analyze theoretically the relation between the objective functions of the Metric GAN and WGAN, and provide an interpretation for the reason why Metric GAN outperforms WGAN.
\item The TSEGAN method provides a flexible framework, which allows other types of TasNet to be used for the generator in this framework. 
\end{itemize}

The remainder of the paper is organized as follows. Section \ref{sec:TSEGAN} presents our proposed  TSEGAN in detail. Section \ref{sec:Discuss}  theoretically analyzes the relation between Metric GAN and WGAN. Section \ref{sec:Experiments} shows the experimental results. Section \ref{sec:conclusion} gives the conclusion.
\section{Time-domain Speech Enhancement GAN}
\label{sec:TSEGAN}
\begin{figure*}[htbp!]
    \centering
    \setlength{\belowcaptionskip}{1pt}
	\setlength{\abovecaptionskip}{1pt}
	\includegraphics[width=0.9\textwidth]{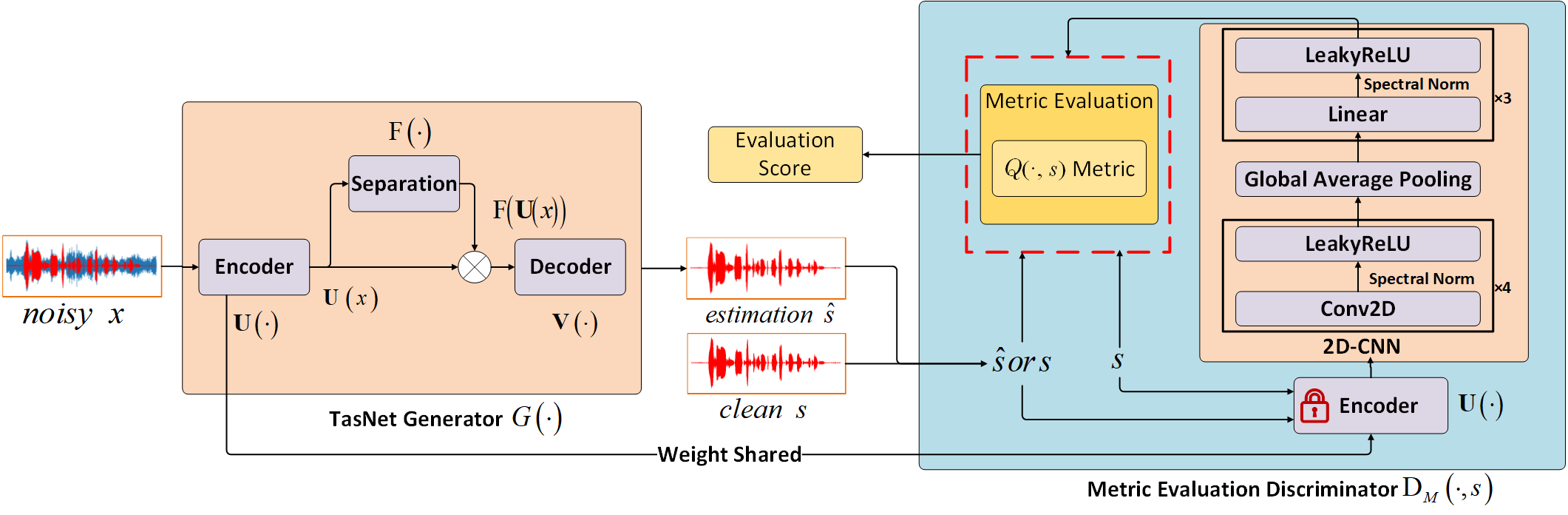}
	\caption{Framework of the proposed TSEGAN, which uses SI-SNR or signal-to-noise ratio (SNR) to calculate the evaluation metric $Q\left( \cdot,s \right) \in \left[ -1,1 \right]$. 
	A reference signal (i.e., $s$) and a signal to be evaluated (i.e., $s$ or $\hat{s}$) are the inputs of $Q\left( \cdot,s \right)$ and the metric discriminator ${{\operatorname{D}}_{M}}\left( \cdot,s  \right)$. 
	The encoder and decoder are denoted as $\mathbf{U}\left( \cdot  \right)$ and $\mathbf{V}\left( \cdot  \right)$, respectively. The separation function $\operatorname{F}\left( \cdot  \right)$ is applied to the latent feature representation $\mathbf{U}\left( x \right)$ to obtain the mask $\operatorname{F}\left( \mathbf{U}\left( x \right) \right)$. Note that, $\mathbf{U}\left( \cdot \right)$ is fixed during training the discriminator, and will be updated during the training for the generator. The output of ${{\operatorname{D}}_{M}}\left( \cdot,s  \right)$ is a score indicating the quality of the speech (i.e., $s$ or $\hat{s}$).}
\label{fig:1}
\end{figure*}

Our proposed TSEGAN uses TasNet as the generator, and we design a metric evaluation discriminator, which improves the 
training stability of TasNet and facilitates the theoretical analysis in Section \ref{sec:Discuss}.
The framework of our proposed TSEGAN is shown in Figure \ref{fig:1}. 

\subsection{TasNet Generator}
\label{ssec:Conv-TasNet generator}
Consider the noisy speech ${x}\left( t \right) \!=\! {s}\left( t \right) \!+\! {n}\left( t \right) \!\in\! {{\mathbb{R}}^{\mathit{T} }}$, where ${s}\left( t \right) \!\in\! {{\mathbb{R}}^{ \mathit{T} }}$ is the clean speech that needs to be estimated, and ${n}\left( t \right) \!\in\! {{\mathbb{R}}^{ \mathit{T}}}$ is noise, $t$ is the discrete time index, and $T$ is the number of samples. Our aim is to estimate the clean speech $s(t)$. The Conv-TasNet method \citep{7} achieves this by using TasNet and the SI-SNR loss, defined as follows: 
\begin{equation}
\label{eq:si-snr}
    {{\text{SI-SNR}} \left( \hat{s},s \right)} = 10\log_{10}{ \left( {\left \| \alpha {s} \right \|}_2^{2} / {\left \| \alpha {s} - {\hat s} \right \|}_2^{2} \right) },
\end{equation}
where $\alpha = \left \langle {s}, {\hat s} \right \rangle / {\left \| {s} \right \|}_2^{2} $, and ${\hat s}$ denotes the enhanced speech.
In the SI-SNR loss, the scale factor $\alpha$ can be affected by the signal dynamics, which may result in unstable training and performance degradation \citep{le2019sdr}. It was shown in \cite{21} that using the SI-SNR loss can lead to uncontrolled level changes in the enhanced signal, as a result, additional scaling may be required for automatic speech recognition (ASR).

In our TSEGAN, we use TasNet \citep{7} as the generator
${{\operatorname{G}}}\left( \cdot  \right)$, which replaces the STFT and iSTFT in the T-F domain with an encoder $\mathbf{U}\left( \cdot  \right)$ and a decoder $\mathbf{V}\left( \cdot  \right)$ respectively.
Here, the encoder is used to transform the noisy ${x}$ into a representation $\mathbf{U}\left( {x} \right)$ in a latent space, and the decoder aims to transform it back to the enhanced speech signal ${\hat s}$.
A function $\operatorname{F}\left( \cdot  \right)$ is applied to $\mathbf{U}\left( x \right)$ to calculate the latent feature mask, which can be described as follows:
\begin{equation}
\label{eq:3}
 \hat{s}=\mathbf{V}\left( \mathbf{U}\left( {x} \right) \odot \operatorname{F}\left( \mathbf{U}\left( {x} \right) \right) \right),
\end{equation}
where $ \odot $ denotes Hadamard product. 
%$\operatorname{F}\left( \cdot  \right)$  is the separation model used to calculate the latent feature mask. 
With $\mathbf{U}\left( {x} \right)$ and its mask $\operatorname{F}\left( \mathbf{U}\left( {x} \right) \right)$, we can obtain the enhanced speech ${\hat s}$ using Eq. \eqref{eq:3}. Here, we use  an ${L}_{p}$ norm constraint ${\left \| {\hat{s}} - s \right \|}_{p}^{p}$ on ${{\operatorname{G}}}\left( \cdot  \right)$ \citep{Zhang2020}, and denote the enhanced speech as ${\hat{s}} = {{\operatorname{G}}}\left( x  \right)$. Then,  ${{\operatorname{G}}}\left( \cdot  \right)$ can be optimized by a metric evaluation discriminator, which can mitigate the issue of the scaling problem associated with ${\hat s}$, as discussed next.

\subsection{Metric Evaluation Discriminator}
\label{ssec:One dimensional discriminator}
We design a metric evaluation discriminator to 
mitigate the  waveform scaling issue of the SI-SNR loss,
where an evaluation metric $ Q\left( \cdot,s \right)$ is introduced to guide the discriminator optimization and calculate the loss of the discriminator.

Our discriminator consists of a 2D-CNN module, a locked encoder that shares the weight with the encoder in the generator, and a metric evaluation module, as shown in Figure \ref{fig:1}. 

In our work, $Q\left( \cdot, s \right)$ is defined as follows:
\begin{equation}
\label{eq:Q}
Q\left( \cdot, s \right) = tanh \left( \text{SI-SNR} \left( \cdot,s \right) / \beta \right),
\end{equation}
where $\beta$ is the parameter to adjust the sensitivity interval of $tanh \left( \cdot \right)$, empirically set as 100 in our experiments. Then, we can get the metric $Q\left( \cdot, s \right) \in \left[ -1, 1 \right]$ from $\text{SI-SNR} \left( \cdot,s \right) \in \mathbb{R}$ with $tanh \left( \cdot \right)$ function, where $Q\left( \cdot, s \right) = 1$ represents the best case and $Q\left( \cdot, s \right) = -1 $ the worst case, corresponding to the best and worst SI-SNR value, respectively.

The reason why we define $ Q\left( \cdot, s \right)$ in terms of SI-SNR 
% instead of $Q_{\text{PESQ}}\left( \cdot, s \right)$ or $Q_{\text{STOI}}\left( \cdot, s \right)$ 
is to verify that the metric evaluation can mitigate the problem of arbitrary scaling caused by the SI-SNR loss. 
In our work, $Q\left( \cdot, s \right)$ is a constant value that is used to limit the evaluation score of the discriminator, which is not directly used as a loss function. As a result, we can mitigate the scaling problem by removing the scale factor $\alpha$ in the gradient update step during the training of the discriminator.
Meanwhile, using our $ Q\left( \cdot, s \right)$, the discriminator of TSEGAN can be mapped to the same range (i.e., real number domain $\mathbb{R}$) as that of WGAN. This provides us a way to analyze the relationship between the objective functions of Metric GAN and WGAN. 

Here, the discriminator is used to evaluate the clean speech $s$ and the enhanced speech $\hat{s}$
 with the metric $Q\left(s, s \right)$ and $Q\left(\hat{s}, s \right)$, 
respectively.  
The output of the discriminator is an evaluation score within a continuous range, indicating the quality of the enhanced speech. 
The continuous range of the output means that it is no longer limited to the probability value of binary classification as in \cite{10}. 
As a result, it can improve the enhancement performance guided through the metric evaluation of the discriminator,  thus improving the generalization ability of the speech enhancement model, which is not only applicable for TasNet \citep{7}, but also can be beneficial for other TasNet based methods.

\subsection{Loss Function}
\label{ssec:loss function}
In our TSEGAN framework, we use metric evaluation for adversarial training. For the purpose of performance comparison and theoretical analysis, we also  adopt the WGAN loss for model training.

When using the Wasserstein distance instead of metric evaluation in TSEGAN, namely W-TSEGAN-L1, the discriminator and generator loss functions are given, respectively, as follows:
\begin{equation}
\label{eq:6}
\begin{array}{*{35}{l}}
   {{L}_{\operatorname{D}_{W}}}={{\mathbb{E}}}\left[- {{\operatorname{D}}_{W}}\left( s,s \right) + {{\operatorname{D}}_{W}}\left( \operatorname{G}\left( x \right),s \right) \right]
\end{array}
\end{equation}
\begin{equation}
\label{eq:7}
    {{L}_{\operatorname{G}_{W}}}=-{{\mathbb{E}}}\left[ {{\operatorname{D}}_{W}}\left( \operatorname{G}\left( x \right),s \right) \right]  + {\lambda} {\left \| \operatorname{G}\left( x \right) - s \right \|}_{1} 
\end{equation}
where ${\mathbb{E}}\left[ \cdot \right]$ is the expectation.
${\operatorname{D} _W}\left(  \cdot,s  \right)$ denotes the discriminator of the W-TSEGAN-L1 and $\lambda$ is the penalty parameter for ${L}_{p}$ norm constraint ${\left \| \operatorname{G}\left( x \right) - s \right \|}_{p}^{p}$. Here,  the L1 constraint is used  (i.e., $p=1$) for W-TSEGAN-L1 with $\lambda\!=\!200$. Moreover, another loss without L1 constraint (i.e., $\lambda\!=\!0$) is also adopted for ablation study, namely W-TSEGAN.  The output of ${\operatorname{D} _W}\left(  \cdot ,s \right)$ is in the real number domain ${{\mathbb{R}}}$.

Regarding our method with metric evaluation, the discriminator and generator loss functions are formulated, respectively, as follows:
\begin{equation}
\label{eq:4}
\begin{split}
{{L}_{\operatorname{D}_{M}}}={{\mathbb{E}}}& \left[ {{\left( {{\operatorname{D}}_{M}}\left( s,s \right)-Q\left( s,s \right) \right)}^{2}} + \right.\\ 
&\left. {\left( {{\operatorname{D}}_{M}}\left( \operatorname{G}\left( x \right),s \right)-Q\left( \hat{s},s \right) \right)}^{2} \right]\\ 
\end{split}
\end{equation}
\begin{equation}
\label{eq:5}
{{L}_{\operatorname{G}_{M}}}={{\mathbb{E}}} \left[{{\left( {{\operatorname{D}}_{M}}\left( \operatorname{G}\left( x \right),s \right)-q \right)}^{2}}\right] + {\lambda} {\left \| \operatorname{G}\left( x \right) - s \right \|}_{1} 
\end{equation}
where ${{\operatorname{D}}_{M}}\left( \cdot,s  \right) \in [-1, 1]$ is the discriminator.
%  and we set $Q\left( s,s \right) = 1$ for the best evaluation. 
$q$ is the target evaluation score that we want generator to arrive through the discriminator, which is set as 1 for the best evaluation $Q\left( s,s \right)$.
Here, a model with $\lambda=200$, using $Q\left( \cdot,s \right)$ from SI-SNR is designed, namely M-TSEGAN-L1.
Another two models without norm constraint (i.e., $\lambda=0$), using $Q\left( \cdot,s \right)$ from SI-SNR and SNR are also designed for the purpose of comparison, namely M-TSEGAN-SISNR and M-TSEGAN-SNR, respectively.

It is worth noting that the speech quality of the generator can be constrained by specifying the value of $q$, which means our proposed TSEGAN framework may be used in wider application scenarios (e.g.,  data augmentation).

\section{Analysis of Metric GAN and WGAN}
\label{sec:Discuss}
The study in \cite{16} points out that Metric GAN has better performance than WGAN in terms of empirical evaluations, but without theoretical analysis and interpretation. To our knowledge, there is no existing theoretical analysis to show the relationship between them. Here, we provide an interpretation for the relationship between WGAN and Metric GAN (i.e., W-TSEGAN-L1 and M-TSEGAN-L1 in our framework).

In our M-TSEGAN-L1, the metric $Q\left( \cdot,s \right) \in \left[ -1,1 \right]$ is calculated from SI-SNR, such that we have ${{\operatorname{D}}_{M}}\left( \cdot,s  \right) \in \left[ -1,1 \right]$, mapped from ${{\mathbb{R}}}$.
If we let $Q\left( \cdot,s \right) = \text{SI-SNR} \left( \cdot,s \right)$, Eq. \eqref{eq:4} can be converted to the objective function ${{L}_{\operatorname{D}_{M_{\mathbb{R}}}}}$ with a discriminator ${{\operatorname{D}}_{{{M}'}}}\left( \cdot,s  \right)$ in ${{\mathbb{R}}}$ as follows:
\begin{equation}
\label{eq:9}
\begin{split}
{{L}_{\operatorname{D}_{M_{\mathbb{R}}}}}={{\mathbb{E}}}& \left[ {{\left( {{\operatorname{D}}_{M'}}\left( s,s \right)-{\text{SI-SNR}}\left( s,s \right) \right)}^{2}} + \right.\\ 
&\left. {\left( {{\operatorname{D}}_{M'}}\left( \operatorname{G}\left( x \right),s \right)-{\text{SI-SNR}}\left( \hat{s},s \right) \right)}^{2} \right] \\
\end{split}
\end{equation}
According to the norm equivalence property:
\begin{equation}
\label{eq:norm_eq}
\arg \min_{y} \left \| y \right \|_2^2 \Leftrightarrow \arg \min_{y}\left \| y \right \|_1^1
\end{equation}
we infer that Eq. \eqref{eq:9} is equivalent to Eq. \eqref{eq:L1_d}:
\begin{equation}
\label{eq:L1_d}
\begin{split}
L_{\operatorname{D}_{M_{\mathbb{R}}}}' = \mathbb{E} &\left[ \left| \operatorname{D}_{M'}\left( s,s \right) -\text{SI-SNR}\left( s,s \right) \right| + \right. \\
&\left. \left| \operatorname{D}_{M'}\left( \operatorname{G}\left( x \right),s  \right) - \text{SI-SNR}\left( \hat{s}, s \right) \right| \right] 
\end{split}
\end{equation}
The SI-SNR of the clean speech $s$ is $\infty$ theoretically. 
Here, we assume the SI-SNR of the clean speech $c\!=\!{\text{SI-SNR}}\left( s,s \right)$ is a value approaching infinity in $L_p$ space, as the upper bound metric value. 
Then, we set $d\!=\!{\text{SI-SNR}}\left( \hat{s},s \right)$, where $d$ is a bounded value. 
According to the Minkowski inequality: 
\begin{equation}
\label{eq:14}
%\small
\begin{split}
    \left \| a \right \|_1 + \left \| b \right \|_1 \ge \left \| a+b \right \|_1 
\end{split}
\end{equation}
Eq. \eqref{eq:L1_d} has the unequal relationship as follows:
\begin{equation}
\label{eq:12}
\begin{split}
   L_{\operatorname{D}_{M_{\mathbb{R}}}}'= {{\mathbb{E}}}& \left[ \left |c-{{ {{\operatorname{D}}_{{{M'}}}}\left( s,s \right) }}\right | + \left |{{ {{\operatorname{D}}_{{{M'}}}}\left( \operatorname{G}\left( x \right),s \right) }}-d \right | \right] \\
    \!\ge\! {{\mathbb{E}}}& \left[ \left | {\left( c-d \right) \!-\! \left( {{\operatorname{D}}_{{{M'}}}} \left( s,s \right)  \!-\!   {\operatorname{D}}_{{{M'}}} \left( \operatorname{G}\left( x \right) ,s \right) \right) }\right | \right]
\end{split}
\end{equation}
According to the Minkowski inequality, in Eq. \eqref{eq:12}, the equal sign holds if and only if $\left( c-{{\operatorname{D}}_{{{M}'}}}\left( s,s \right)\right) =k\left( {{ {{\operatorname{D}}_{{{M}'}}}\left( \operatorname{G}\left( x \right),s \right)-d }} \right)$, where $k \ge 0$. Therefore, there are two conditions with which the equal sign holds, respectively: 
\begin{enumerate}[(i)]
\item $\operatorname{D}_{M'}\left( \operatorname{G}\left( x \right),s \right) \!\ge\! d$, $ \operatorname{D}_{M'}\left( s,s \right) \!\le\! c$.

From this, we can deduce $\left( c-d \right) \!\ge\! \left({{\operatorname{D}}_{{{M}'}}} \left( s,s \right)  \!-\! {\operatorname{D}}_{{{M}'}} \left( \operatorname{G}\left( x \right),s \right)\right)$.
\item $\operatorname{D}_{M'}\left( \operatorname{G}\left( x \right),s \right) \!<\! d$, $ \operatorname{D}_{M'}\left( s,s \right) \!=\! c$.

From this, we can deduce $\left( c-d \right) \!<\! \left({{\operatorname{D}}_{{{M}'}}} \left( s,s \right)  \!-\! {\operatorname{D}}_{{{M}'}} \left( \operatorname{G}\left( x \right),s \right)\right)$.
\end{enumerate}
For simplification, we define:
\begin{equation}
\label{eq:13}
{{L}_{\operatorname{D}_{M}'}}\!=\! {{\mathbb{E}}} \left[ \left | {\left( c-d \right) \!-\! \left( {{\operatorname{D}}_{{{M'}}}} \left( s,s \right)  \!-\!   {\operatorname{D}}_{{{M'}}} \left( \operatorname{G}\left( x \right),s \right) \right) }\right | \right]
\end{equation}
Eq. \eqref{eq:13} is the objective function of ${{\operatorname{D}}_{{{M}'}}}\left( \cdot,s  \right)$. 
With the first condition (i),
during optimization, it is expected to have $\left({{\operatorname{D}}_{{{M}'}}} \left( s,s \right)  \!-\! {\operatorname{D}}_{{{M}'}} \left( \operatorname{G}\left( x \right),s \right)\right) \!\longrightarrow^-\! \left( c-d \right)$. Therefore, %when $\left( c-d \right)$ is large enough, 
when $\left( c-d \right) \!\ge\! \left({{\operatorname{D}}_{{{M}'}}} \left( s,s \right)  \!-\! {\operatorname{D}}_{{{M}'}} \left( \operatorname{G}\left( x \right),s \right)\right)$, 
optimizing Eq. \eqref{eq:13} is equivalent to optimize Eq. \eqref{eq:10}:
\begin{equation}
\label{eq:10}
\begin{split}
  {{L}_{\operatorname{D}_{M}'}}\Leftrightarrow {{\mathbb{E}}}\left[- {{\operatorname{D}}_{{{M}'}}}\left( s,s \right) + {{\operatorname{D}}_{{{M}'}}}\left( \operatorname{G}\left( x \right),s \right) \right]
\end{split}
\end{equation}
Note that, with the second condition (ii), we expect to have $\left({{\operatorname{D}}_{{{M}'}}} \left( s,s \right)  \!-\! {\operatorname{D}}_{{{M}'}} \left( \operatorname{G}\left( x \right),s \right)\right) \!\longrightarrow^+\! \left( c-d \right)$, which is not equivalent to the optimization of Eq. \eqref{eq:10}. 
However, this drives that the evaluation score of the discriminator will approach the actual quality of the speech.

For the generator $\operatorname{G}\left( \cdot \right)$, its target metric score $q'$ is $c\!=\!{\text{SI-SNR}}\left( s,s \right)$. Similar to the process from Eq. \eqref{eq:13} to Eq. \eqref{eq:10}, ${{\operatorname{D}}_{{{M}'}}}\left( \operatorname{G}\left( x \right),s \right)$ is expected to approach $c$, which leads to large ${{\operatorname{D}}_{{{M}'}}}\left( \operatorname{G}\left( x \right),s \right)$. Then, the Eq. \eqref{eq:5} is equivalent to Eq. \eqref{eq:11} as follows:
\begin{equation}
\label{eq:11}
    {{L}_{\operatorname{G}_{M}}}\Leftrightarrow -{{\mathbb{E}}}\left[ {{\operatorname{D}}_{{{M}'}}}\left( \operatorname{G}\left( x \right),s \right) \right] + {\lambda} {\left \| \operatorname{G}\left( x \right) - s \right \|}_{p}^{p} 
\end{equation}
Now we can find that Eq. \eqref{eq:10} and Eq. \eqref{eq:11} are equivalent to the objective functions of W-TSEGAN-L1 (i.e., Eq. \eqref{eq:6} and Eq. \eqref{eq:7}), respectively. 
When the equal sign in Eq. \eqref{eq:12} holds, the continuous indicator of ${{\operatorname{D}}_{{{M}'}}}\left( \cdot,s \right)$ will be equivalent to the Wasserstein distance between the clean speech $s$ and its estimation $\hat{s}$. 
This means that 
the solution space of Metric GAN is larger than that of WGAN, and 
the solution space of WGAN is included in that of Metric GAN.

There is a difference between WGAN and Metric GAN in the optimization of the discriminator. In WGAN, the Wasserstein distance between $s$ and $\hat{s}$ is maximized, and as a consequence, the optimization of the discriminator may limit the improvement of the generator. However, in Metric GAN, the discriminator follows a differentiable metric evaluation model. Each enhanced speech has its own metric evaluation score. The improvement of the generator will not be affected by the discriminator, which can be seen as an adapted metric evaluation loss function, that is helpful for the training of the generator.
The experimental analysis will be provided in Section \ref{ssec:analysis}.

\section{Experiments}
\label{sec:Experiments}

\subsection{Dataset}
\label{ssec:dataset}
We choose the Voice-Bank + DEMAND (VBD) dataset \citep{18,19} including 11,572 and 824 noisy-clean speech pairs for training and testing, respectively. There are 30 speakers in the VBD dataset, where 28 speakers are used for training, and 2 speakers are used for testing. 
{The training set includes 40 different conditions \citep{18}, including 10 types of noise (i.e., 2 artificial and 8 from the Demand database \citep{19}) and 4 SNR settings (i.e., 15dB, 10dB, 5dB, and 0 dB). Whereas the test set includes 20 different conditions \citep{18}, including 5 types of noise (all from the Demand database) and 4 SNR settings (i.e., 17.5dB, 12.5dB, 7.5dB, and 2.5 dB). Note that, the test set is totally unseen by the training set.}
We set the sampling rate of the VBD dataset to 16kHz \citep{9} and divide the training set into segments with each of 1 second.
{In addition, we randomly split 5\% of the training set as the validation set, and the remainder is used as the development set for models training. The validation set is used to show the performance of model during the training process and decide whether the training process should be stopped.}
\subsection{Experimental Setup}
\label{ssec:set up}
To build the TasNet model as the generator, we set window length $L$ as 2 ms, other hyper-parameters are set according to \cite{7}. 
For comparison, we set Conv-TasNet using the SI-SNR loss as a baseline as in \cite{7}, and the TasNet structure using the MSE loss, namely, Conv-TasNet-MSE as another baseline. 

In addition, to  show the effectiveness of our  TSEGAN framework, and demonstrate that the use of metric evaluation based loss is superior to the Wasserstein distance,  we conduct experiments using TSEGANs, i.e., W-TSEGAN, W-TSEGAN-L1, M-TSEGAN-L1, M-TSEGAN-SNR and M-TSEGAN-SISNR, as compared with other speech enhancement methods, i.e., SEGAN and Conv-TasNets.  

All TSEGANs are constrained to 1-Lipschitz continuous by spectral normalization in discriminator to make the training stable \citep{20}. 
The 2D-CNN module includes 4 convolutional layers and 3 fully connected layers, with the kernel sizes of 5, 7, 9 and 11 respectively, and the strides are set as 2.
Adam optimizer \citep{25} is employed for all model training, with the initial learning rate set as  0.001, which is halved if the {SI-SNR value} is not improved on validation split in three consecutive epochs. The batch size is set as 20 for all the models.
{The early stopping strategy is also employed in our model training. If the SI-SNR value is not improved in ten consecutive epochs on validation split, the training stage will be stopped.}
\subsection{Performance Analysis}
\label{ssec:evaluation}
We apply objective evaluation criterion to demonstrate the effectiveness of our proposed TSEGAN, where perceptual evaluation of speech quality (PESQ) (from -0.5 to 4.5) \citep{rix2001perceptual}, composite measure (i.e., CSIG, CBAK, COVL in \cite{hu2007evaluation}), Segmental SNR (SSNR) \citep{hansen1998effective} and SI-SNR \citep{luo2018tasnet} are employed  as the performance metrics.% to evaluate the quality of the enhanced speech signals. 

Table \ref{tab:ablation} shows the experimental results  for speech enhancement. From the table, we can see  that all methods based on Conv-TasNet and our TSEGAN are better than SEGAN in terms of PESQ and SI-SNR. Conv-TasNet-MSE outperforms Conv-TasNet in terms of all the perceptual evaluation metrics. Especially, the SSNR of the baseline is negative, which confirms the degraded performance caused by the SI-SNR loss, whereas all our proposed models can significantly improve the SSNR performance. 
\begin{table*}[htbp]
    % \vspace{-0.8cm}  %调整与上文的垂直距离
	\setlength{\belowcaptionskip}{1pt}
	\setlength{\abovecaptionskip}{1pt}
	\centering
	\caption{Objective metrics performance of the proposed TSEGAN on VBD dataset}
	  \small{
	  \begin{tabular}{ccccccccc}
	  \toprule
	  \multicolumn{2}{c}{Methods} & Loss function & PESQ & CSIG & CBAK & COVL & SI-SNR & SSNR \\
	  \midrule
	  \multicolumn{2}{c}{Noisy} & - & 1.97 & 3.35 & 2.44 & 2.63 & 8.64 & 4.62\\
	  \midrule
	  \multicolumn{2}{c}{SEGAN} & LSGAN+L1 &  2.16 & 3.48 & 2.94 & 2.80 & 5.06 & 7.73 \\
	  \multicolumn{2}{c}{Conv-TasNet (baseline)} & SI-SNR & 2.19 & 3.41 & 1.93 & 2.79 & 15.37 & -8.91 \\
	  \multicolumn{2}{c}{Conv-TasNet-MSE} &  MSE & 2.54 & 3.77 & 3.28 & 3.15 & 18.87 & 9.84 \\
	  \midrule
	  \multicolumn{2}{c}{W-TSEGAN} & WGAN & 2.40 & 3.56 & 3.15 & 2.96 & 18.28 & 9.05 \\
      \multicolumn{2}{c}{W-TSEGAN-L1} & WGAN+L1 & 2.48 & 3.64 & 3.23 & 3.04 & 18.87 & 9.76 \\
      \multicolumn{2}{c}{M-TSEGAN-SNR} & Metric GAN & 2.45 & 3.69 & 3.19 & 3.05 & 18.64 & 9.46 \\
      \multicolumn{2}{c}{M-TSEGAN-SISNR} & Metric GAN &  2.49 & 3.73 & 3.23 & 3.10 & 18.72 & 9.59 \\
	  \multicolumn{2}{c}{M-TSEGAN-L1} &  Metric GAN+L1 &  2.52 & 3.67 & 3.27 & 3.08 & \textbf{19.26} & \textbf{10.16} \\
	  \bottomrule
	  \bottomrule
	  \end{tabular}%
	  }
\label{tab:ablation}%
\end{table*}%
{Regarding our TSEGANs,  from Table \ref{tab:ablation}, we can see that without L1 constraint, the TSEGANs using metric evaluation (i.e., M-TSEGAN-SNR, M-TSEGAN-SISNR) outperform the TSEGAN using Wasserstein distance (i.e., W-TSEGAN) in terms of PESQ, composite measure, SI-SNR and SSNR. In contrast, when L1 constraint is applied, the TSEGAN using metric evaluation (i.e., M-TSEGAN-L1) can outperform the TSEGAN using Wasserstein distance (i.e., W-TSEGAN-L1) in terms of PESQ, composite measure, SI-SNR and SSNR. These comparisons confirm that Metric GAN is better than WGAN. 

The results also show that M-TSEGAN-SISNR can achieve better performance than  M-TSEGAN-SNR, in terms of all metrics, which indicates the metric evaluation calculated from SI-SNR is a better choice for TSEGAN. In addition, when compare the TSEGANs with L1 constraint and the TSEGANs without L1 constraint, we can see that L1 constraint can provide performance improvement. }

When compared to Conv-TasNet-MSE, M-TSEGAN-L1 gives better performance improvement in terms of SI-SNR and SSNR, i.e., with 0.39 dB and 0.32 dB improvements, respectively. This may be because of the use of  $ Q\left( \cdot, s \right)$ with SI-SNR in our TSEGAN. The results show that the metric evaluation process makes our TSEGAN model tend to give better  performance in terms of SNR metric.
Meanwhile, the PESQ and composite measure of Conv-TasNet-MSE and M-TSEGAN-L1 are almost the same.
This may be because we have not found the optimal $\lambda$ for norm constraint. 
In addition, all of our TSEGANs can avoid the scaling problem by using metric evaluation to update the gradients. Therefore, TSEGANs can be considered as a special loss that has excellent generalization ability.
Note that, other TasNet based structures  also can be used as the generator of TSEGAN.
\begin{figure*}[htb]
    \centering
	\setlength{\abovecaptionskip}{1pt}
    \subfigure[]{
        \vspace{-0.8cm}  
        \setlength{\abovecaptionskip}{-0.2cm}   
        \includegraphics[width=0.45\textwidth]{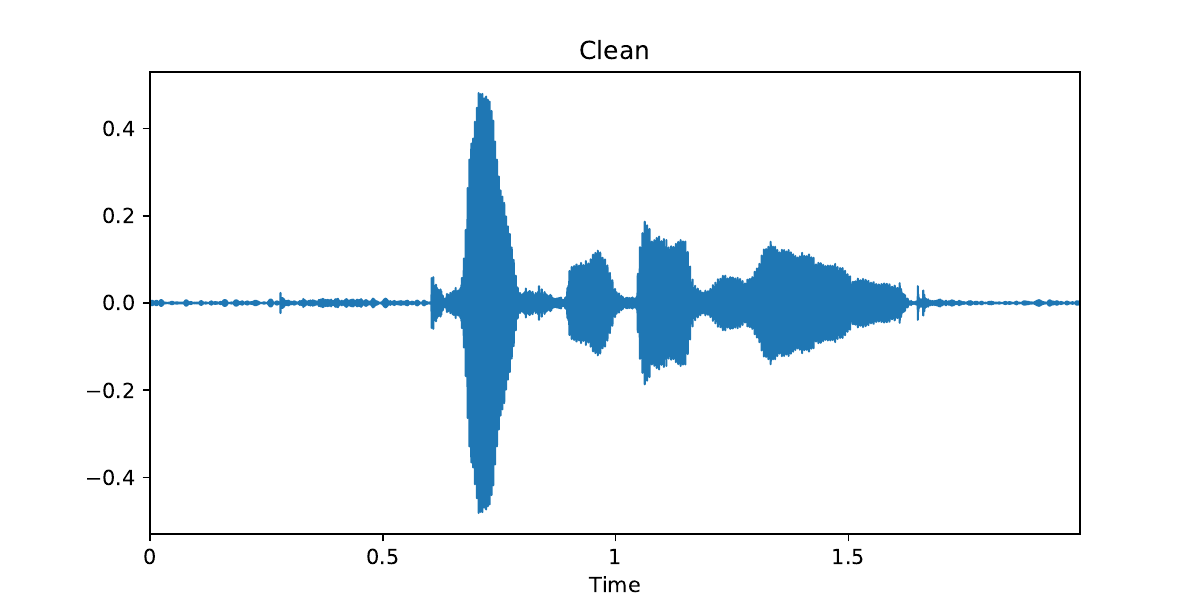}
    }
    \quad
    \subfigure[]{
        \vspace{-0.8cm}  
        \setlength{\abovecaptionskip}{-0.2cm}   
        \includegraphics[width=0.45\textwidth]{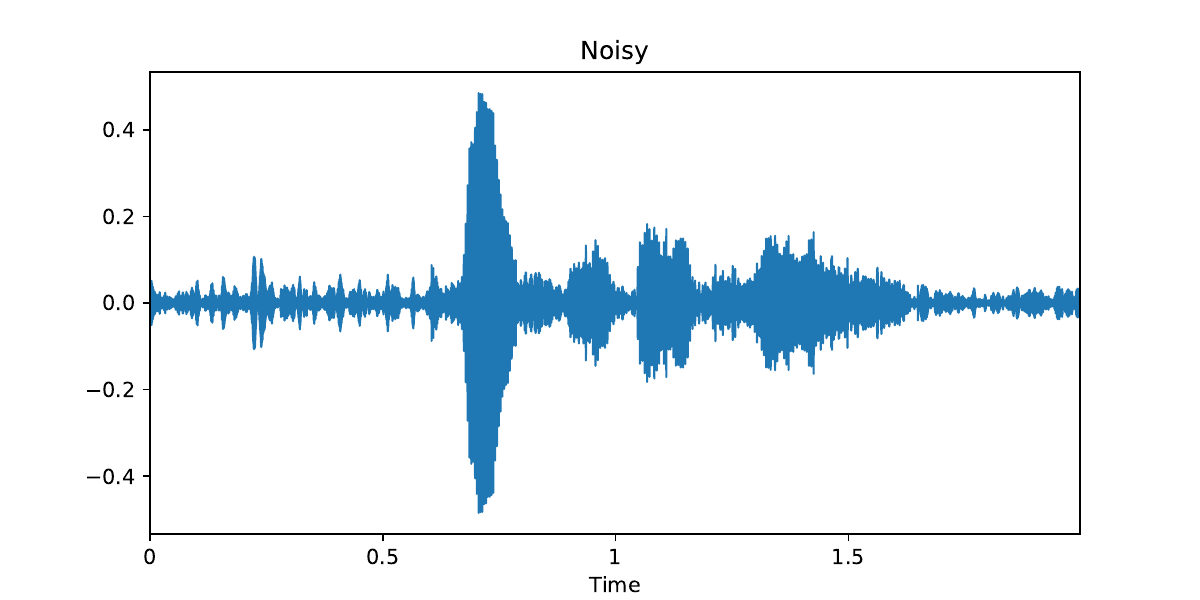}
    }
    \subfigure[]{
        \vspace{-0.8cm}  
        \setlength{\abovecaptionskip}{-0.2cm}
        \includegraphics[width=0.45\textwidth]{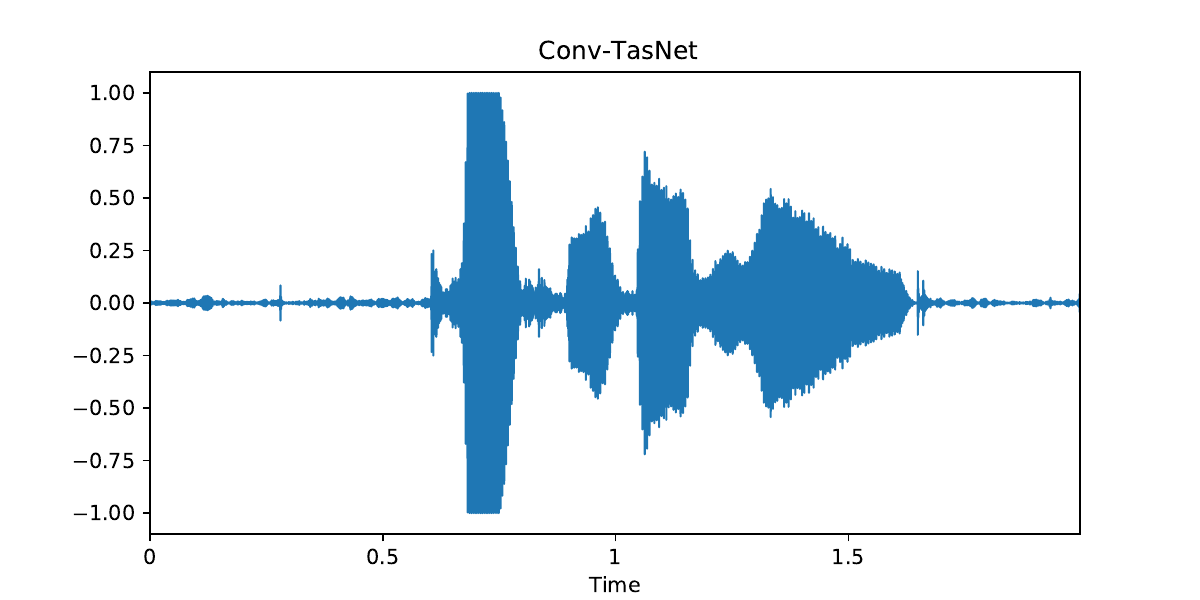}
        \label{sub:3}
    }
    \quad
    \subfigure[]{
        \vspace{-0.8cm}  
        \setlength{\abovecaptionskip}{-0.2cm}
        \includegraphics[width=0.45\textwidth]{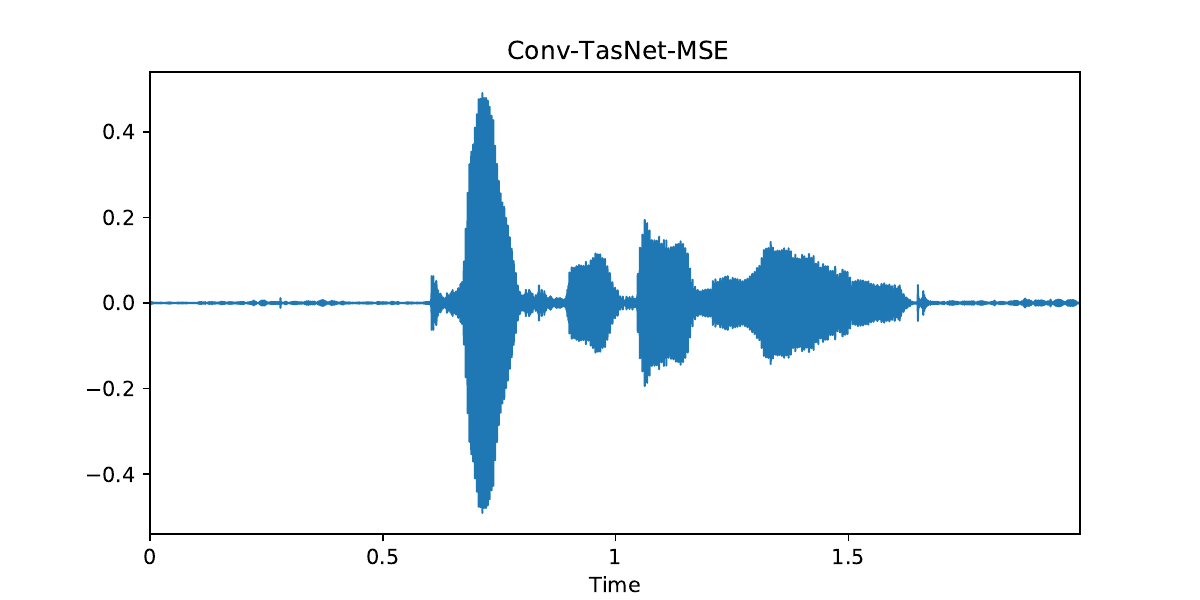}
    }
    \quad
    \subfigure[]{
        \vspace{-0.8cm}  
        \setlength{\abovecaptionskip}{-0.2cm}
        \includegraphics[width=0.45\textwidth]{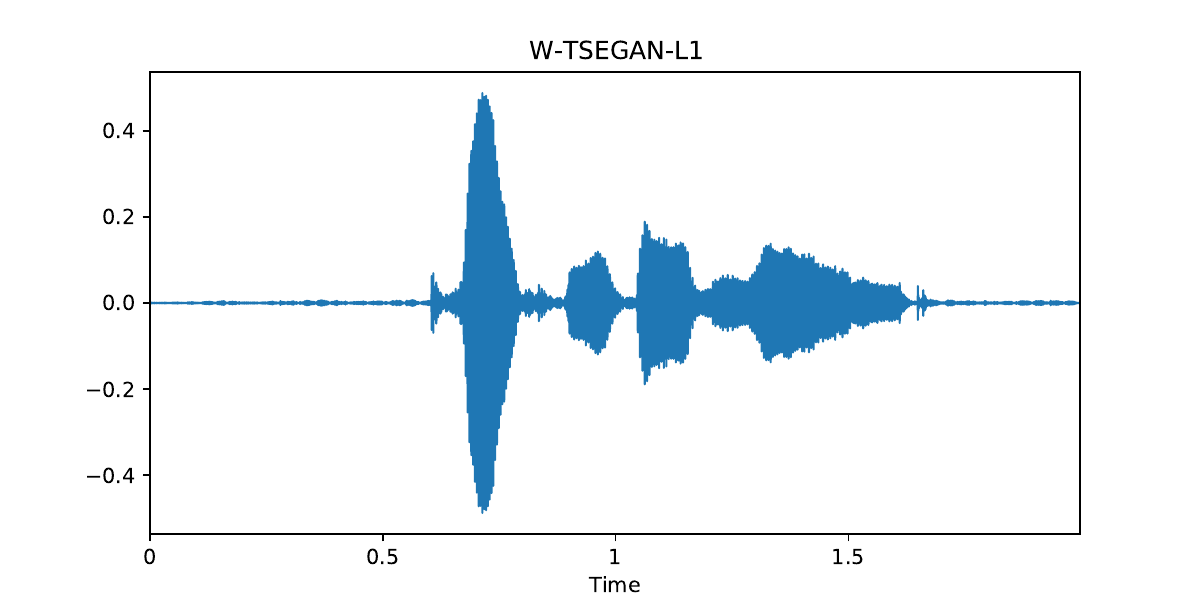}
    }
    \quad
    \subfigure[]{
        \vspace{-0.8cm}  
        \setlength{\abovecaptionskip}{-0.2cm}
        \includegraphics[width=0.45\textwidth]{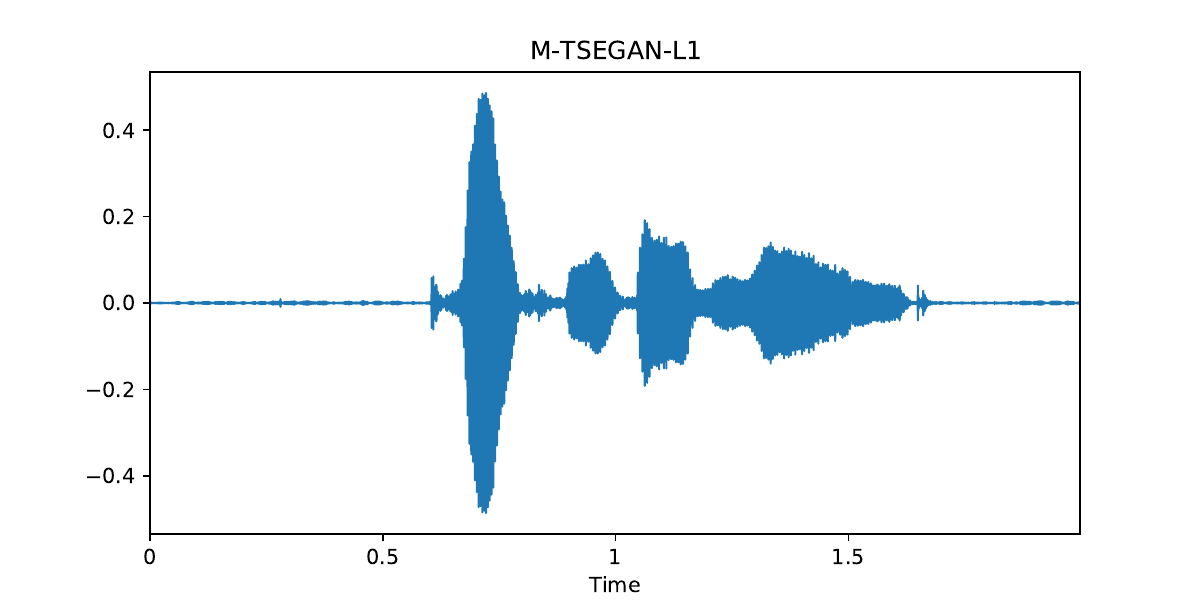}
    }
    \caption{Performance illustration in waveform, where $p257\_342.wav$ in the test set is used for illustration.}
\label{fig:2}
\end{figure*}
We also give the enhancement results of different models (i.e., Conv-TasNet, Conv-TasNet-MSE, W-TSEGAN-L1 and M-TSEGAN-L1) in waveform, where a signal (i.e., $p257\_342.wav$)\footnote{Demo files   for more results will be provided along with the source code.} from test set is selected for illustration,  as shown in Figure \ref{fig:2}. Note that, as the test signal from VBD already contains clean signal and its corresponding noisy signal, hence we don't need to add additional noise to the clean speech.
According to  Figure \ref{fig:2}, we can see that the waveform enhanced by Conv-TasNet, i.e., Figure \ref{fig:2} (c), is excessively scaled and remains some noise signal, leading to poor metric performance in Table \ref{tab:ablation}. This phenomenon confirms the reason why Conv-TasNet underperforms in speech enhancement is the arbitrary scaling caused by the SI-SNR loss function. 

\begin{figure*}[htb]
    \centering
    \subfigure[]{
        \vspace{-0.8cm}  
        \setlength{\abovecaptionskip}{-0.2cm}   
        \includegraphics[width=0.3\textwidth]{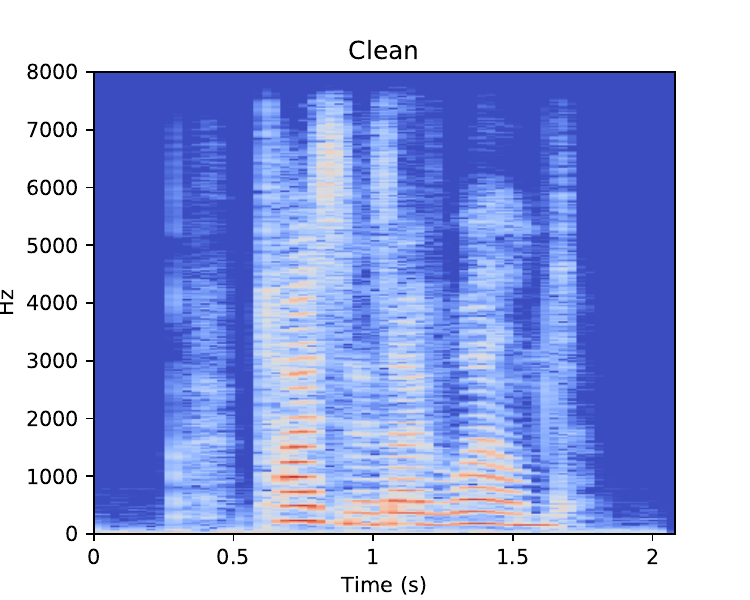}
    }
    \hspace{-0.6cm}
    \quad
    \subfigure[]{
        \vspace{-0.8cm}  
        \setlength{\abovecaptionskip}{-0.2cm}   
        \includegraphics[width=0.3\textwidth]{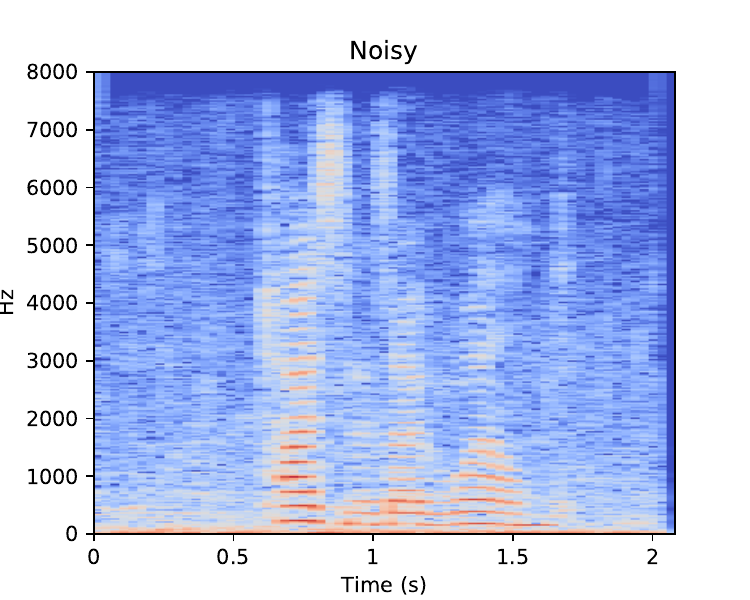}
    }
    \hspace{-0.6cm}
    \quad
    \subfigure[]{
        \vspace{-0.8cm}  
        \setlength{\abovecaptionskip}{-0.2cm}
        \includegraphics[width=0.3\textwidth]{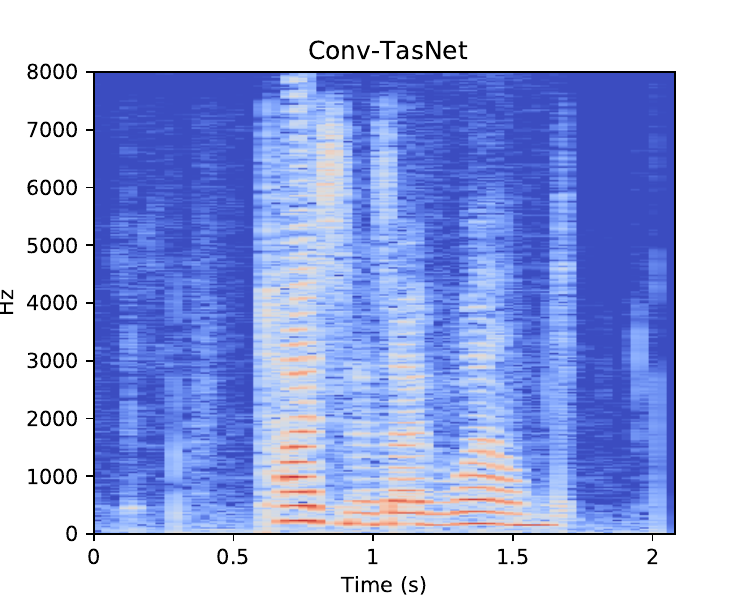}
    }
    \hspace{-0.6cm}
    \quad
    \subfigure[]{
        \vspace{-0.8cm}  
        \setlength{\abovecaptionskip}{-0.2cm}
        \includegraphics[width=0.3\textwidth]{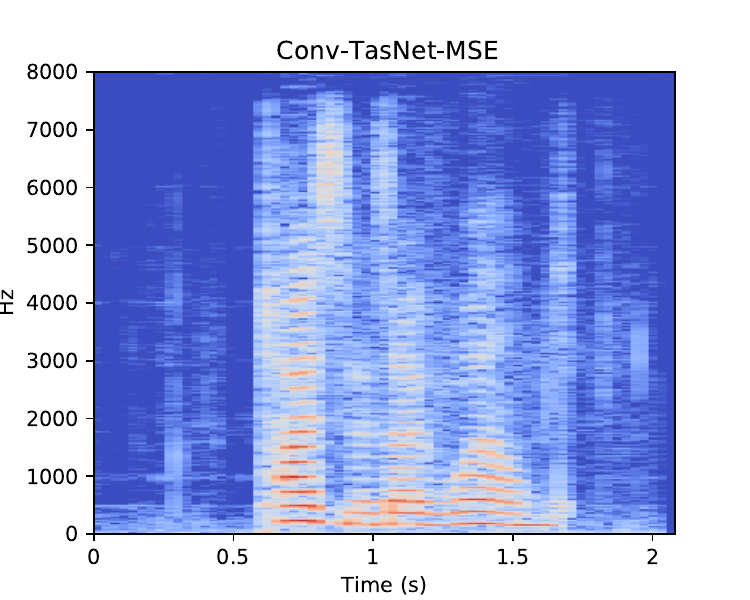}
    }
    \hspace{-0.6cm}
    \quad
    \subfigure[]{
        \vspace{-0.8cm}  
        \setlength{\abovecaptionskip}{-0.2cm}
        \includegraphics[width=0.3\textwidth]{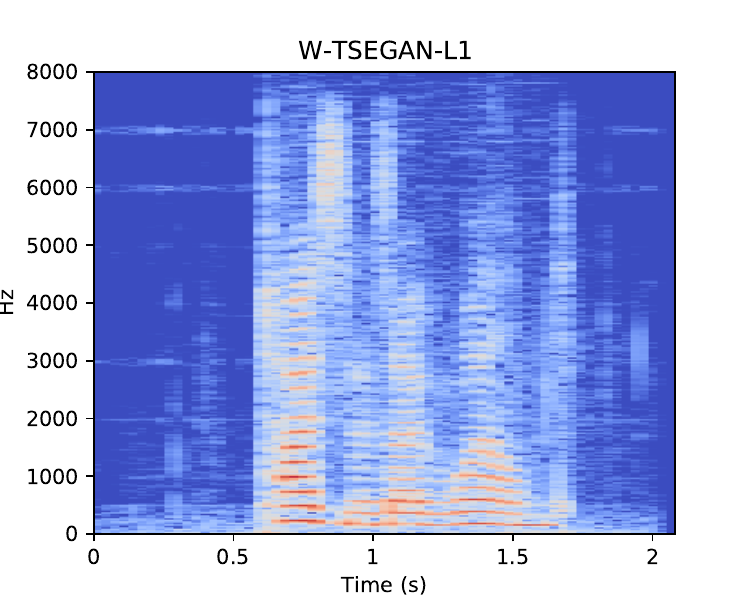}
    }
    \hspace{-0.6cm}
    \quad
    \subfigure[]{
        \vspace{-0.8cm}  
        \setlength{\abovecaptionskip}{-0.2cm}
        \includegraphics[width=0.3\textwidth]{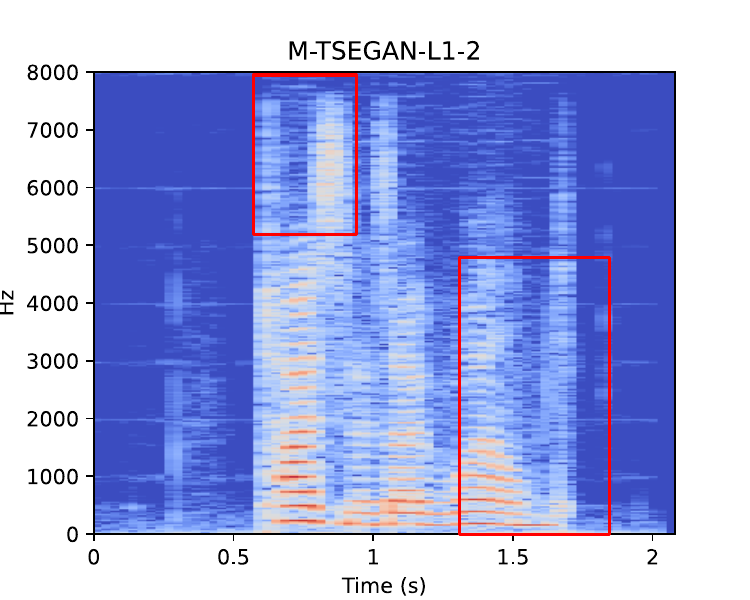}
    }
    \caption{Performance illustration in spectrograms, where $p257\_342.wav$ in the test set is used for illustration. }
\label{fig:3}
\end{figure*}

Meanwhile, in Figure \ref{fig:2}, the waveforms enhanced by Conv-TasNet, W-TSEGAN-L1, M-TSEGAN-L1 all approximate to the clean speech waveform, which means all these models can eliminate noise effectively. We also give the corresponding spectrograms of these waveforms for the purpose of further analysis, as shown in Figure \ref{fig:3}.

{From Figure \ref{fig:3}, we can see that  Figure \ref{fig:3} (d), (e) and (f) have the approximate spectrograms, which confirms the objective metrics as shown in Table \ref{tab:ablation}.  
Though the enhancement results  of Conv-TasNet-MSE, W-TSEGAN-L1, and M-TSEGAN-L1  all remain some noise signal in high frequency band, as shown in Figure \ref{fig:3} (d), (e), and (f) respectively,  the result of M-TSEGAN-L1 has less noise signal in high frequency band. In addition, M-TSEGAN-L1 can also recover the speech components with clear structure, as shown in the red box of Figure \ref{fig:3} (f). This also confirms the performance of Metric GAN is better than WGAN.
On the other hand, the red box in  Figure \ref{fig:3} (f) also shows that M-TSEGAN-L1 can effectively remove  noise signal and provide the enhanced speech with high SNR value.}
\subsection{Relation Analysis between Metric GAN and WGAN}
\label{ssec:analysis}
In this section, we analyze the relation between Metric GAN and WGAN. 
First, we define a discriminator gap estimation $\mathcal{L}_{est}$ as shown in Eq. \eqref{eq:W_estimation}.
\begin{equation}
\label{eq:W_estimation}
    \mathcal{L}_{est}  =  {{\mathbb{E}}}\left[- {{\operatorname{D}}}\left( s,s \right) + {{\operatorname{D}}}\left( \operatorname{G}\left( x \right),s \right) \right]
\end{equation}
where ${{\operatorname{D}}}\left( \cdot, s \right)$ is used in a general sense of the discriminators of Metric GAN (i.e., M-TSEGAN) and WGAN (i.e., W-TSEGAN).

Specifically, for WGAN, $\mathcal{L}_{est} $ is the Wasserstein distance estimation $\mathcal{L}_{W_{est}} $ that is used to optimize the discriminator. Regarding Metric GAN, though it does not require such a distance for discriminator optimization, we  still can obtain such a discriminator gap estimation as described in  Eq. \eqref{eq:W_estimation} during the training of the discriminator, which is named as  $\mathcal{L}_{M_{est}} $ in this work. Therefore, we can use the estimations (i.e., $\mathcal{L}_{W_{est}} $ and $\mathcal{L}_{M_{est}} $) to explore the relation between WGAN and Metric GAN.  

In our experiments,  we additionally calculate and record the $\mathcal{L}_{est} $ values for both WGAN and Metric GAN in  each training step, such that we can get a set of curves for the entire training process to reflect the trend of $\mathcal{L}_{M_{est}} $ and $\mathcal{L}_{W_{est}} $ values. Then we can compare the curves to show the relation between WGAN and Metric GAN.
\begin{figure*}[htb]
    \centering
    \includegraphics[width=0.95\textwidth]{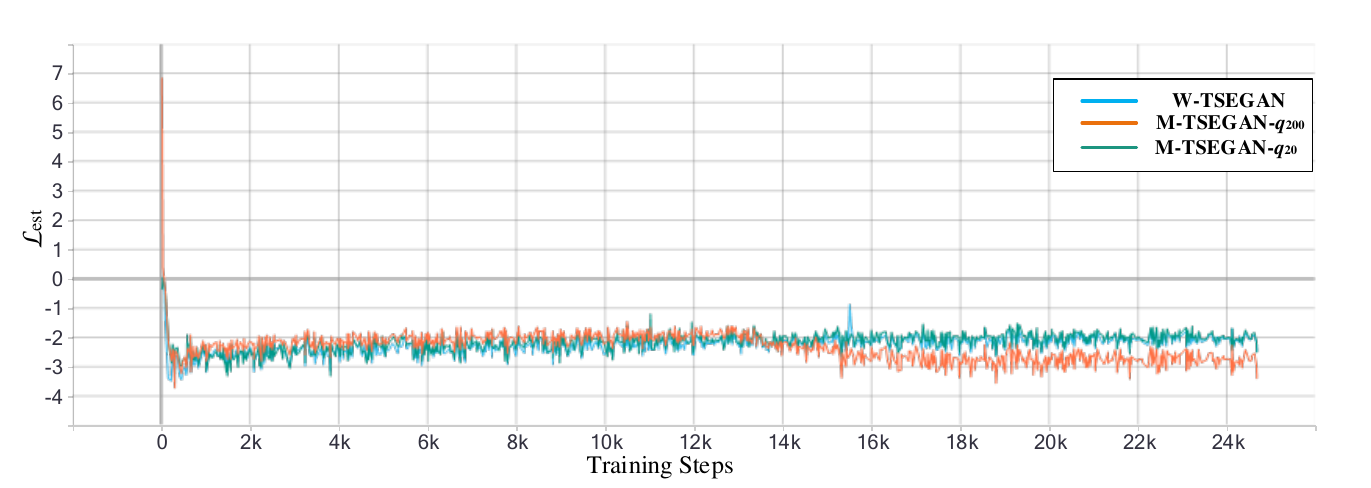}
     \caption{Discriminator gap estimation for Metric GAN and WGAN, where the $\mathcal{L}_{W_{est}} $ curve of W-TSEGAN, and the $\mathcal{L}_{M_{est}}$ curves of M-TSEGAN-$q_{200}$ and M-TSEGAN-$q_{20}$ are showed and compared.}
\label{fig:4}
\end{figure*}

Here, for convenience, we simplify the generator of our TSEGANs with a simpler separation module, i.e., a temporal convolutional network (TCN) with an 8-layers stack. 
In particular, the discriminator loss function of the M-TSEGANs in this section is set according to Eq. \eqref{eq:9} (i.e., $Q\left( \cdot,s \right) = \text{SI-SNR} \left( \cdot,s \right)$), which means the range of ${{\operatorname{D}}}\left( \cdot,  s \right)$ is $\left( -\infty, \infty \right)$ theoretically. 
However, as the infinity value is unable to calculate the gradient in the implementation, the target evaluation score $q$ (i.e., $Q \left( s, s \right)$) of M-TSEGANs will be set to a high constant SI-SNR value to narrow the {${{\operatorname{D}}}\left( \cdot, s \right)$} range in the implementation {(i.e., in the range of $\left[ -q, q \right]$)}. 
Meanwhile, in order to see the influence of $q$ values for the estimation, we set the target metric evaluation score $q$ as 200 and 20, then the M-TSEGANs  can be denoted as  M-TSEGAN-$q_{200}$ and M-TSEGAN-$q_{20}$, respectively.
Figure \ref{fig:4} shows the $\mathcal{L}_{est} $ curves of W-TSEGAN, M-TSEGAN-$q_{200}$ and M-TSEGAN-$q_{20}$ in the training stage.

According to Figure \ref{fig:4}, the $\mathcal{L}_{W_{est}} $ curve of W-TSEGAN and the $\mathcal{L}_{M_{est}} $ curve of M-TSEGAN-$q_{20}$ almost overlap, which means W-TSEGAN and M-TSEGAN-$q_{20}$ have the equivalent $\mathcal{L}_{est}$ trend. This result confirms our analysis in Section \ref{sec:Discuss}: With certain conditions, Metric GAN is equivalent to WGAN. 
However, we can still find that when $q$ value is set as 200, the  $\mathcal{L}_{M_{est}} $ curve (i.e., M-TSEGAN-$q_{200}$) is approximate to that of W-TSEGAN from 0 to 14k steps, but slightly lower than  the $\mathcal{L}_{W_{est}} $ curve after 14k steps.

\begin{figure*}[htb]
    \centering
    \subfigure[M-TSEGAN-$q_{200}$]{
        \vspace{-0.8cm}  
        \setlength{\abovecaptionskip}{-0.2cm}
        \includegraphics[width=0.85\textwidth]{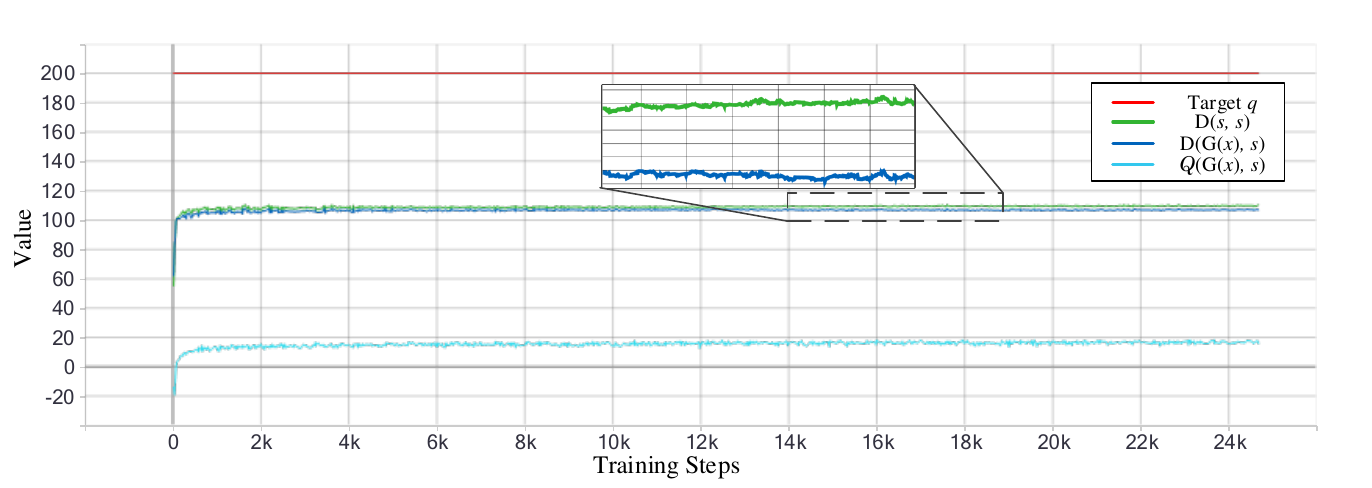}
    }
    \subfigure[M-TSEGAN-$q_{20}$]{
        \vspace{-0.8cm}  
        \setlength{\abovecaptionskip}{-0.2cm}
        \includegraphics[width=0.85\textwidth]{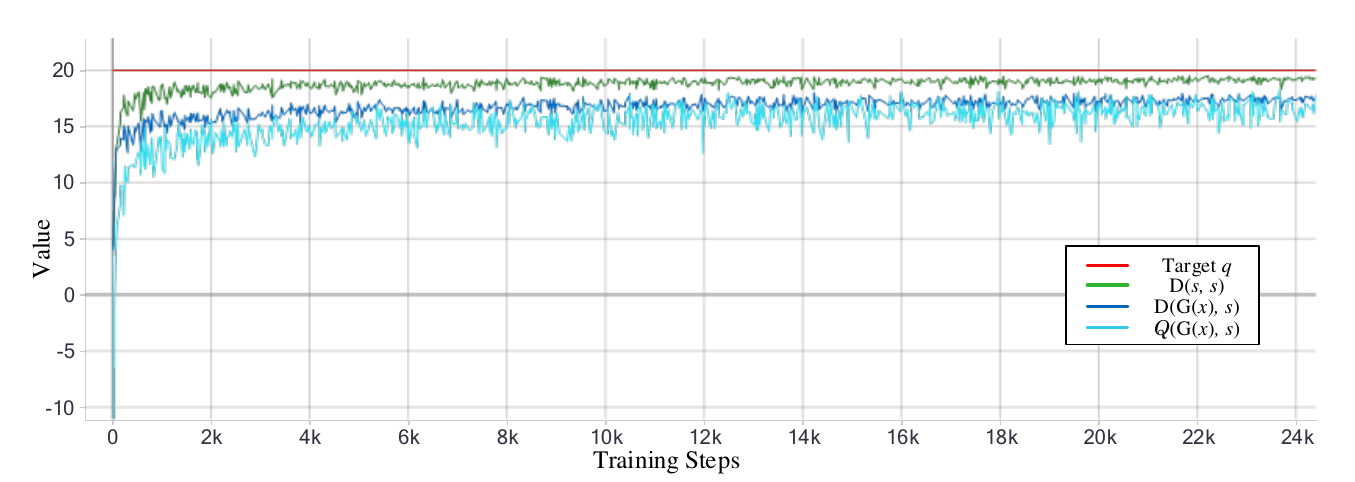}
    }
    \caption{The curves of ${{\operatorname{D}}}\left( \cdot, s \right)$ and $Q \left( \cdot, s \right)$ of M-TSEGANs with different target $q$ values. Here, (a) is the set of curves of M-TSEGAN-$q_{200}$, and  (b) is the set of curves of  M-TSEGAN-$q_{20}$. }
\label{fig:5}
\end{figure*}
In order to confirm the specific conditions for the equivalence and find the reason of this diverse curves trend, 
we record the values of ${{\operatorname{D}}}\left( s, s \right)$, ${{\operatorname{D}}}\left( \operatorname{G}\left( x \right), s \right)$, target evaluation score $q$ (i.e., $Q \left( s, s \right)$) and $Q \left( \operatorname{G}\left( x \right), s \right)$ (i.e., from Eq. \eqref{eq:9}) of M-TSEGAN-$q_{200}$ and M-TSEGAN-$q_{20}$ in each training step, and form two sets of curves of ${{\operatorname{D}}}\left( \cdot, s \right)$ and $Q \left( \cdot, s \right)$. These two sets of curves are drawn in Figure \ref{fig:5}, corresponding to M-TSEGAN-$q_{200}$ and M-TSEGAN-$q_{20}$ respectively. 
Then, we compare the ${{\operatorname{D}}}\left( \cdot, s \right)$ curves and $Q \left( \cdot, s \right)$ curves in each subfigure of Figure \ref{fig:5}, to show the relation between ${{\operatorname{D}}}\left( \cdot, s \right)$ and $Q \left( \cdot, s \right)$ and analyze the condition and the reason.

In Figure \ref{fig:5}, target $q$ is the $Q \left( s, s \right)$, which is the SI-SNR value of the clean speech (i.e., $c$ in Eq. \eqref{eq:13}).  $Q \left( \operatorname{G}\left( x \right), s \right)$ denotes the SI-SNR value of the enhanced speech (i.e., $d$ in Eq. \eqref{eq:13}). 
As can be seen from Figure \ref{fig:5}, the M-TSEGAN-$q_{200}$ curves always satisfy the condition: $\operatorname{D}\left( \operatorname{G}\left( x \right),s \right) \!\ge\! d$ and $ \operatorname{D}\left( s,s \right) \!\le\! c$, and the curves of M-TSEGAN-$q_{20}$ satisfy the condition $\operatorname{D}\left( \operatorname{G}\left( x \right),s \right) \!\ge\! d$ and $ \operatorname{D}\left( s,s \right) \!\le\! c$ in most time steps. 
This is corresponded with the condition (i) we gave in Section \ref{sec:Discuss} when Metric GAN is equivalent to WGAN. Therefore,   the results can be used to support our discussion about the relation between Metric GAN and WGAN.

Regarding the  difference between the curves of M-TSEGAN-$q_{200}$ and W-TSEGAN in Figure \ref{fig:4}, the reason is that, in the M-TSEGAN-$q_{200}$ training stage, ${{\operatorname{D}}}\left( s,s \right)$ is expected to approach higher $q$, whereas  ${{\operatorname{D}}}\left( \operatorname{G}\left( x \right),s \right)$ is expected to approach the corresponding SI-SNR value $Q \left( \operatorname{G}\left( x \right), s \right)$, according to Eq. \eqref{eq:9}. This opposite tendency increases the distance between ${{\operatorname{D}}}\left( s,s \right)$ and ${{\operatorname{D}}}\left( \operatorname{G}\left( x \right),s \right)$ as shown in Figure \ref{fig:5} (a), making the $\mathcal{L}_{M_{est}}$ of M-TSEGAN-$q_{200}$ lower than the $\mathcal{L}_{W_{est}}$ of W-TSEGAN.

\begin{figure*}[htb]
    \centering
    \includegraphics[width=0.95\textwidth]{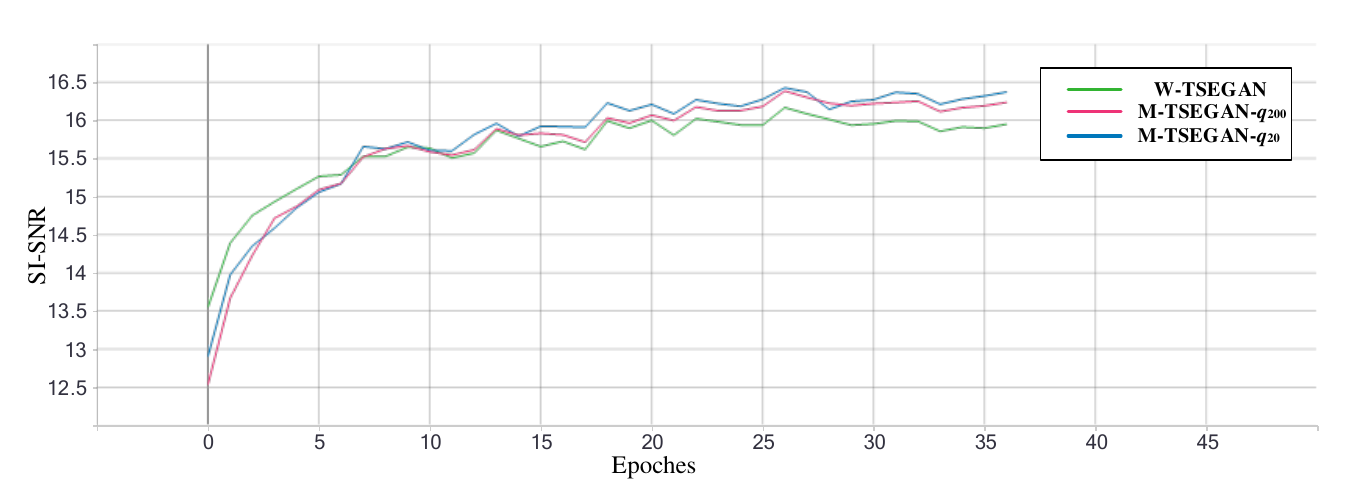}
    \caption{The SI-SNR performance  on the validation set of W-TSEGAN, M-TSEGAN-$q_{200}$ and M-TSEGAN-$q_{20}$. }
\label{fig:6}
\end{figure*}

In order to further  analyze Metric GAN and WGAN, we give the SI-SNR performance curves on the validation set as shown in Figure \ref{fig:6}. 
Though $\mathcal{L}_{est}$ curves are approximate to each other (as shown in Figure \ref{fig:4}), we can observe that M-TSEGANs can still achieve better SI-SNR performance on the validation set as shown in Figure \ref{fig:6}. 
This is an interesting phenomenon which demonstrates Metric GAN is better than WGAN.
It shows that the use of metric evaluation in M-TSEGANs can improve the performance of the metric  corresponding to $Q \left( \cdot, s \right)$ (i.e., SI-SNR), even if the equivalence condition is satisfied. 

Moreover, we can see M-TSEGAN-$q_{20}$  outperforms  M-TSEGAN-$q_{200}$ as shown in Figure \ref{fig:6}.  The reason is that, for  M-TSEGAN-$q_{200}$, the distance between  ${{\operatorname{D}}}\left( \cdot, s \right)$ and $Q \left( \cdot, s \right)$ is too large (as shown in Figure \ref{fig:5}), such that the  metric evaluation can not fit the metric $Q \left( \cdot, s \right)$ very well. In contrast, with a smaller $q$, the metric evaluation of M-TSEGAN-$q_{20}$ can better fit the corresponding $Q \left( \cdot, s \right)$  than  M-TSEGAN-$q_{200}$. We will further study the influence of $q$ value in our future work. 
\section{Conclusion}
\label{sec:conclusion}
We have presented a novel time-domain speech enhancement framework with generative adversarial network, which employs metric evaluation to optimize the generator and mitigate scale ambiguity of the waveform caused by the SI-SNR loss. Experimental results show the effectiveness of our proposed TSEGAN.  
Moreover, we provided an interpretation for the relation between the Metric GAN and WGAN, and illustrated the advantage of Metric GAN.
According our analysis, under certain conditions, the WGAN can be seen as a special case of the Metric GAN, which provides an interpretation why the Metric GAN outperforms the WGAN. 
In our future work, we will explore the use of other GANs based on this study.

\bibliography{refs}

\end{document}